\documentclass[reqno,12pt]{article}
\usepackage{amsmath,amsfonts,amssymb,amsthm,amstext,amscd,eucal,xcolor}
\usepackage[all]{xy}
\usepackage{hyperref}
\usepackage{epsfig}
\usepackage{color}
\usepackage{graphicx}
\usepackage[active]{srcltx}
\makeatletter \@addtoreset{equation}{section}

\makeatletter\renewcommand\section{\@startsection {section}{1}{\z@}%
                                   {-3.5ex \@plus -1ex \@minus -.2ex}
                                   {2.3ex \@plus.2ex}%
                                   {\normalfont\large\bfseries}}
\renewcommand\subsection{\@startsection{subsection}{2}{\z@}%
                                     {-3.25ex\@plus -1ex \@minus -.2ex}%
                                     {1.5ex \@plus .2ex}%
                                     {\normalfont\bfseries}}

\newcommand{\be}{\begin{equation}}
\newcommand{\ee}{\end{equation}}
\newcommand{\bea}{\begin{eqnarray}}
\newcommand{\eea}{\end{eqnarray}}
\newcommand{\bse}{\begin{subequations}}
\newcommand{\ese}{\end{subequations}}
\newcommand{\bi}{\begin{itemize}}
\newcommand{\ei}{\end{itemize}}

\newcommand{\beq}{\begin{eqnarray}}
\newcommand{\eeq}{\end{eqnarray}}

\newcommand{\nn}{\nonumber}

\def\s2s1{S$^2\times$S$^1$ }
\def\Label#1{\label{#1}%
  \smash{\hbox to0pt{\raise1ex\hbox{\tiny[#1]}\hss}}}
\def\noLabels{\let\Label=\label}
\def\nobbibitem{\let\bbibitem=\bibitem}

\begin{document}
\baselineskip 18pt%
\begin{titlepage}
\vspace*{20mm}
\begin{center}
{\Large{\textbf{Universality of Area Product: Solutions with Conical Singularity}}}
\vspace*{8mm}

Hanif Golchin\footnote{h.golchin@uk.ac.ir} \\
\vspace*{0.4cm}
{ \it Faculty of Physics, Shahid Bahonar University of Kerman, \\
PO Box 76175, Kerman, Iran}\\
\vspace*{1.5cm}
\end{center}

\begin{abstract}
It has been observed that the area product of horizons for many black hole solutions is mass independent and satisfy the universality relation $A_+A_-=(8\pi)^2 N$, where $N$ is related to the quantized charges of the solution as angular momentum and electric charge. In this work the same analysis is done for black hole and black ring solutions with conical singularity. We find that the area product is still mass independent and regardless of the horizon topology, the conical characteristic ($\kappa$) of the solutions, appears in the universality relation as $\kappa A_+A_-=(8\pi)^2 N$\,. We also check that the first law of black hole inner mechanics is satisfied for these solutions.
\end{abstract}

\end{titlepage}

\addtocontents{toc}{\protect\setcounter{tocdepth}{2}}
\tableofcontents

\section{Introduction}
Within the context of string theory, it has been suggested \cite{Horowitz:1996fn}-\cite{Visser:2012zi} that the area of black hole event horizon might be quantized as
\be
A=8\pi \ell_p^2(\sqrt{N_1}+\sqrt{N_2})\,, \quad N_1, N_2\in \mathbb{N}\,.
\ee
In the case of black hole solutions that possess both inner and outer horizons, there is also a string theory-inspired conjecture for the area product \cite{Cvetic:1997uw}-\cite{Castro:2012av}
\be
A_+A_-=(8\pi \ell_P^2)^2\,N\,, \quad N\in \mathbb{N}\,.
\ee
In recent years the area product of multi-horizon black holes is received many attentions \cite{Cvetic:2010mn}-\cite{Anacleto:2013esa}. It has been observed  that the product of horizon areas in many solutions, is independent of the mass and it depends only to the quantized charges as angular momentum and electric charge. This property is called ``universality'' of area product \cite{Cvetic:2010mn,Ansorg:2009yi,Ansorg:2010ru}. It is also shown \cite{Cvetic:2010mn,Castro:2012av} that this universality can be generalized to the higher dimensional black hole and black ring solutions. Black rings \cite{Emparan:2001wn,Emparan:2008eg} are solutions with the horizon topology of $S^1\times S^{d-3}$, where $d\ge 5$ is the space-time dimension.

It has been shown in \cite{Chen:2012mh} that by considering just the physical inner and outer horizons\footnote{There are negative or complex roots for $g^{rr}=0$. In \cite{Chen:2012mh} the biggest and the smallest positive real roots are considered as outer and inner horizons respectively.}, the entropy product $S_+S_-$ is mass independent when $T_+S_+=T_-S_-$\,, where $T_{\pm}$ are hawking temperature of the inner and outer horizons. It is observed \cite{Chen:2012mh} that for some solutions such as the Myers-Perry black holes in more than 5 dimensions, $T_+S_+\neq T_-S_-$ so the entropy product $S_+S_-$ is mass dependent. However it is shown in \cite{Cvetic:2010mn} that by considering all the horizons, the entropy product becomes mass independent. Thermodynamics of black hole (ring) horizons can also be related to the CFT duals of black hole (ring) \cite{Guica:2008mu}-\cite{Sadeghian:2015hja}, this relation is studied in \cite{Chen:2012mh,Chen:2012yd}.

The universality of area product is investigated for some other solutions: in the case of black holes in higher curvature gravity, it is observed \cite{Castro:2013pqa} that the universality of area product fails in general. For the black hole solutions containing the NUT charge, it is also shown in \cite{Xu:2015mna,Debnath:2015tda} that the area product is not universal. However the mass independence of area product is observed for the regular black hole \cite{Pradhan:2015wnl} and for the Acoustic Black Holes \cite{Anacleto:2013esa}. 

This paper is organized as follows. In section \ref{s2} we review, very briefly, some examples for the universality of area product from both black hole and ring solutions. In section \ref{s3} after introducing the conical characteristic of a solution, we concentrate on the area product for some black hole and ring solutions that contain conical singularity. We also verify the first law of thermodynamics and the Smarr relation for the inner horizon of these solutions. Section \ref{s4} is devoted to the concluding remarks.
\section{Universality of area product} \label{s2}
In this section we review some examples for the mass independence of the area product.  hereafter throughout the paper we set the Newton's gravitational constant $G=1$.

{\bf The Kerr-Newman black hole}\\ This four dimensional solution \cite{Newman:1965my} is characterized by mass ($M$), spin ($J$) and electric charge ($Q$). It has been shown \cite{Ansorg:2009yi} that the area product of inner and outer horizons for this solution is universal as
\be \label{unikn}
A_+A_-=16\pi^2(4J^2+Q^4)\,.
\ee
In four and five dimensions, asymptotically flat charged and rotating black hole solutions  can be characterized by six parameters \cite{Cvetic:1996kv,Cvetic:1996xz}. In four dimensions, these parameters are $M$, $J$ and four charges $Q_i,\,i=1,2,3,4$ and in five dimensions the black hole possess $M$, two angular momentum $J_{\phi}, J_{\psi}$ and three charges $Q_1, Q_2, Q_3$. It is shown \cite{Cvetic:2010mn,Chen:2012mh} that the area product in four and five dimensions respectively goes to
\be
A_+A_-=64\pi^2\bigg(J^2+\prod^{4}_{i=1}Q_i\bigg),\quad A_+A_-=64\pi^2\bigg(J_{\phi}J_{\psi}+\prod^{3}_{i=1}Q_i\bigg),
\ee
which are universal explicitly.

{\bf Double rotating black ring}\\
 In the case of asymptotically flat black solutions, the area product is universal regardless of the topology of the horizons.  For a 5 dimensional double rotating neutral black ring (the horizon topology of $S^1\times S^2$) \cite{Pomeransky:2006bd}, it has been checked \cite{Castro:2012av} that the area product is universal as
 \be
 A_+A_-=64\pi^2J_{\phi}^2\,,
 \ee
where the $S^1$ circle of the black ring lies at $\psi$ direction and $J_{\phi}$ is the angular momentum perpendicular to it.

{\bf Single rotating dipole black ring}\\
The universality of area product is also observed for the  single spin dipole black ring \cite{dipole}. This solution possesses dipole charge $q$ in addition to its angular momentum $J_{\psi}$ along the $S^1$ of the ring. It has been shown for this solution that \cite{Castro:2012av,Chen:2012yd}
\be
A_+A_-=64\pi^2J_{\psi}q^3\,,
\ee
which means the universality for the dipole black ring.

\section{Universality and inner black hole mechanics for solutions with conical singularity} \label{s3}
In this section we investigate the area product for black solutions that contain the conical singularity. First we introduce the conical characteristic of these solutions.
\subsection{The conical characteristic}
In general, a solution may contain conical singularity in the space-time. In this case there is a deficit (or excess) for azimuthal angle: $\Delta \phi=2\pi/ \kappa$\,. In the other words by $\kappa>1$ ($\kappa<1$), one obtains conical deficit (excess) in the space-time, while the space-time is regular for $\kappa=1$\,. In order to find the $\kappa$ factor on the horizon, one may expand the $x-\phi$ part\footnote{$x$ denotes the polar coordinate.} of the horizon metric around $x=x_0$ where the $\phi \phi$ component of the horizon metric vanishes. This part takes to the form
\be \label{hmexp}
ds^2_H=\frac{A}{x-x_0}dx^2+B(x-x_0)d\phi^2,
\ee
where $x_0=1, -1$ are the north and south poles on the horizon metric respectively. Now using the transformation $x-x_0=cr^2$, one can rewrite the above part as
\be
ds^2_H=4cA(dr^2+\kappa_{\pm}^2r^2d\phi^2), \quad \kappa_{\pm}^2=\frac{B}{4A}\,.
\ee
$\kappa_+,\kappa_-$ denote the $\kappa$ factor on the north  and south poles on the event horizon, respectively. It is always possible to rescale the $\phi$ coordinate in a manner that $\kappa_-=1$ or $\kappa_+=1$. This means that the horizon is a ``distorted'' sphere which is consisting of a regular southern hemisphere (if we set $\kappa_-=1$) that joined to a conic space in the north pole. The conical characteristic of the solution is given by 
\be \label{k}
\kappa=\frac{\kappa_+}{\kappa_-}\,. 
\ee

In this section we concentrate on the black solutions with conical singularity. We see that the area product depends  to the quantized charges and to the conical characteristic of the solutions, but it is independent of mass. In the other words, the area product is universal for these solutions. We also check the first law of black hole thermodynamics for these solutions in the following.

\subsection{The charged rotating C-metric} 
The C-metric \cite{kw} which describes a pair of uniformly accelerating black holes can be generalized to the black holes containing rotation and charge (the Kerr-Newman black hole) \cite{Hong:2004dm,Griffiths:2005qp,Astorino:2016ybm}. The charged rotating C-metric solution is in the form \cite{Astorino:2016ybm}
\bea \label{qacm}
&ds^2 &\!\!= \frac{1}{(1+ A x r)^2} \Big[ \frac{f(r) + a^2 h (x)}{r^2 + a^2 x^2 } dt^2  - \frac{r^2 + a^2 x^2 }{f(r)} dr^2 + \frac{r^2 + a^2 x^2 }{h(x)} dx^2 \\
&+&\!\!\!\!\!\! \frac{a^2(1-x^2)^2 f(r) + (a^2+r^2)^2 h(x)}{r^2+a^2 x^2} \Delta_\phi^2\, d\phi^2  + 2 \frac{a (1-x^2) f(r) + a (a^2+r^2) h(x) }{r^2+a^2 x^2} \Delta_\phi dt d\phi  \Big],\nn
\eea
where 
\bea
f(r) &=& (A^2 r^2 -1) (r-r_+) (r-r_-),\nn\\
h(x)&=&(1 - x^2) (1+ A x r_+) (1+ A x r_-),\nn\\
\Delta_\phi&=&\frac{1}{1+2mA+A^2(q^2+a^2)}\,.
\eea
In the above solution there are four parameters $m, A, q$ and $a$ which are related to the mass, acceleration, electric charge and angular momentum of the solution, respectively. Ranges of the coordinates are $-1\leq x\leq 1$, $0\leq \phi\leq 2\pi$ and the gauge field in this solution is given by
\be
 A_\mu = \left[- \frac{q r}{r^2 + a^2 x^2} , 0 , 0 , - \frac{ - q r a ( 1-x^2 )}{r^2 + a^2 x^2}    \right]\,,
 \ee
It is also shown in \cite{Hong:2004dm} that the inner and outer horizons are given by
\be
 r_\pm = m \pm \sqrt{m^2-q^2-a^2}\,.
 \ee
For this solution, the area of outer and inner horizons are
\be \label{carea}
A_{\pm}=\!\int\!\! \sqrt{\gamma}\,\Big|_{r=r_{\pm}}\!\!\!\!\!\! dx d\phi=\!\int\!\!  \sqrt{g_{xx}g_{\phi \phi} }\,\Big|_{r=r_{\pm}}\!\!\!\!\!\! dx d\phi=\frac{4\pi \Delta_\phi(r_{\pm}^2 + a^2)}{1-A^2 r_{\pm}^2}\,,
\ee
where $\gamma$ is determinant of the induced line element obtained by setting $dt=dr=0$\,. Rewriting the metric (\ref{qacm}) in ADM form $ds^2=-N^2 dt^2+g_{\phi\phi}\big(d\phi+N^\phi dt\big)^2+g_{rr}dr^2+g_{xx}dx^2$, we calculate temperature on the outer and inner horizons
\be \label{ctemp}
T_+=\frac{\left(N^2\right)'}{4\pi \sqrt{g_{rr}N^2}}\bigg|_{r=r_+}\!\!\!\!\!\!=\frac{(r_+-r_-)\left(1-A^2r_+^2\right)}{4\pi\left(r_+^2+a^2\right)}\,,\qquad T_-=\frac{(r_+-r_-)\left(1-A^2r_-^2\right)}{4\pi\left(r_-^2+a^2\right)}\,.
\ee
Now it is easy to check that
\be
T_+S_+=T_-S_-\,,
\ee
where $S_{\pm}=A_{\pm}/4$ is the entropy. This means that the area product for (\ref{qacm}) is universal.  The electric charge, angular momentum and mass of the solution can be obtained as
\bea
&&Q=\frac{1}{8\pi} \int \!\epsilon_{\alpha \beta \mu \nu} F^{\mu\nu} dx d\phi=q\Delta_{\phi}\,, \qquad J=\frac{1}{16\pi} \int \!\epsilon_{\alpha \beta \mu \nu} \nabla^{\mu}\phi^{\nu}dx d\phi=am\Delta_{\phi}^2\,,\nn\\
&&M=-\frac{1}{8\pi} \int \!\epsilon_{\alpha \beta \mu \nu} \nabla^{\mu}\xi^{\nu}dx d\phi=m\Delta_{\phi}\,,
\eea
where $F_{\mu\nu}=\partial_{\mu} A_{\nu}-\partial_{\nu} A_{\mu}$\,,\, $\phi^{\nu}$ and $\xi^{\nu}$ are the rotational and timelike Killing fields $\partial_{\phi}$\, and $\partial_t$ respectively. Doing the same steps as (\ref{hmexp})-(\ref{k}) for metric (\ref{qacm}) one can find the conical characteristic of this solution as
\be
\kappa=\frac {1-2mA+A^2 ({q}^{2}+{a}^{2})}{1+2mA+{A}^{
2} ({q}^{2}+{a}^{2})}\,.
\ee
Noting these results, one can check that the universality of the area product in this case takes to the form
\be
\kappa A_+A_-=16\pi^2(4J^2+Q^4)\,,
\ee
which is similar to the area product of the Kerr-Newman black hole (\ref{unikn}), up to the appearing of conical characteristic $\kappa$.

The electric potential and angular velocity on the inner and outer horizon can also be find as
\be 
\Phi_{\pm}=-\chi^{\mu}A_{\mu}\Big|_{r=r_{\pm}}\!\!\!\!=\frac{qr_{\pm}}{r_{\pm}^2+a^2}\,, \qquad \Omega_{\pm}=N^{\phi}\Big|_{r=r_{\pm}}\!\!\!\!=\frac{a}{(r_{\pm}^2+a^2)\Delta_{\phi}}\,.
\ee
Now one can check that the first law of thermodynamics, $dM=T_+ dS_+ +\Omega_+\, dJ+\Phi_{\!+}dQ$, is satisfied. It is easy to check that the Smarr relation and the first law of thermodynamics for the inner horizon are satisfied too:
\be
dM=-T_- dS_-+\Omega_-\, dJ+\Phi_{\!-}dQ\,, \qquad M=-2T_-S_-+2\Omega_-J+\Phi_-Q\,.
\ee

\subsection{Black ring with rotating $S^2$}
A black ring solution with rotation only along $S^2$, is introduced in \cite{Figueras:2005zp}. Note that in the case of black ring solutions ($S^1\times S^2$ topology), tension of the ring may be canceled by the centrifugal force due to the rotation along $S^1$. So the black ring in \cite{Figueras:2005zp} is unbalanced which means that it contains the conical singularity. The solution is
\bea \label{s2r}
ds^2 &=& -\frac{H(y,x)}{H(x,y)} 
  \left[ dt - 
   \frac{\lambda y R\sqrt{\sigma}(1-x^2)}
        {H(y,x)} \, d\phi\right]^2 
  + \frac{R^2H(x,y)}{(x-y)^2} 
  \Bigg[\frac{dx^2}{(1-x^2)F(x)}
    - \frac{dy^2}{(1-y^2)F(y)}\nn\\
    &-& \frac{(1-y^2) F(x)}{H(x,y)}d\psi^2
   + \frac{(1-x^2)F(y)}{H(y,x)}d\phi^2
   \Bigg]\,,
\eea
where
\be
F(\xi)=1+\lambda \xi+\sigma \xi^2\,, \qquad H(\xi_1,\xi_2)=1+\lambda \xi_1+\sigma (\xi_1\,\xi_2)^2\,.
\ee
Coordinate $\psi$ in this solution, parametrizes the $S^1$ and $(x,\phi)$ show the $S^2$. $x,y$ are ring coordinates that lie in the ranges
\be
-1\leq x\leq 1\,,\qquad -\infty <y\leq -1\,,
\ee
with the asymptotic infinity that is located at $x=y=-1$. The solution contains three parameters $\sigma, R$ and $\lambda$ which $\sigma$ controls the rotation of $S^2$, $R$ is related to the black ring radius and $\lambda$ lies in the range $2\sqrt{\sigma}<\lambda<1+\sigma$. The conical singularity can be removed at $x=y=-1$ by setting the period of $\phi, \psi$ coordinates as
\be
\Delta\phi=\Delta\psi=\frac{2\pi}{\sqrt{1-\lambda+\sigma}}\,.
\ee
However the conical singularity remains at $x=-1$ and similar to (\ref{hmexp})-(\ref{k}), one can find the conical characteristic of this solution as
\be \label{kf}
\kappa_-=1\,,\qquad \kappa=\kappa_+=\sqrt{\frac{1+\lambda+\sigma}{1-\lambda+\sigma}}\,.
\ee
Roots of $F(y)$ in (\ref{s2r}) are correspond to the horizons of this black ring solution. So there are two horizons located at $y_{\pm}=\frac{-\lambda \pm \sqrt{\lambda^2-4\sigma}}{2\sigma}$\,. Similar to the previous subsection, one can find the area of horizon, mass, temperature and angular momentum for this solution as \cite{Figueras:2005zp}
\bea
&&A_+=\frac{8\pi^2R^3\lambda}{(1-\lambda+\sigma)\sqrt{y_+^2-1}}\,,\quad M=\frac{3\pi R^2\lambda}{4(1-\lambda+\sigma)}\,,\nn\\
&&T_+=\frac{\sqrt{(\lambda^2-4\sigma)(y_+^2-1)}}{4\pi \lambda R}\,,\quad J^\phi=\frac{\pi R^3 \lambda \sqrt{\sigma}}{(1-\lambda+\sigma)^{3/2}}\,.
\eea
We also calculated the area, temperature and angular velocity of the inner horizon. the result is
\bea
&&A_-=\frac{8\pi^2R^3\lambda}{(\lambda-1-\sigma)\sqrt{y_-^2-1}}\,,\quad T_-=\frac{\sqrt{(\lambda^2-4\sigma)(y_-^2-1)}}{-4\pi \lambda R}\,,\nn\\
&&\Omega_-^{\phi}=\frac{\left(\lambda+\sqrt {\lambda^2-4\sigma}\right)\! \sqrt{1-\lambda+\sigma}}{-2R\lambda\sqrt{\sigma}}\,.
\eea
The inner horizon quantities as well as the outer horizon ones, satisfy the first law of thermodynamics and the Smarr relation 
\be
dM=-T_- dS_-+\Omega_-^{\phi}\, dJ^{\phi}\,,\quad M=\frac32\left(-T_-S_-+\Omega_-^{\phi} J^{\phi}\right).
\ee
For this solution $T_+S_+=T_-S_-$ is satisfied too and using (\ref{kf}), one can check the universality of area product as
\be
\kappa A_+A_-=64\,\pi^2 {J^{\phi}}^2\,.
\ee
Note again the appearance of the conical characteristic $\kappa$ in the above.

\subsection{The unbalanced Pomeransky-Sen'kov black ring}
A neutral black ring solution in 5 dimensions, characterizes by four parameters relating to its mass, two angular momenta and the ring radius, which is introduced in \cite{Chen:2011jb}. In general this solution is unbalanced and contains the conic singularity. 
The solution is in the form \cite{Chen:2011jb}
\bea \label{ubr}
ds^2\!\!\!\!&=&\!\!\!\!-\frac{H(y,x)}{H(x,y)}\Big[ dt\!- \omega_\phi(x,\!y)d\phi-\omega_\psi(x,\!y)d\psi\Big]^2\!\!\!+\!\frac{F(y,x)}{H(y,x)}\,d\phi^2-2\frac{J(x,y)}{H(y,x)}\,d\psi d\phi-\frac{F(x,y)}{H(y,x)}\,d\psi^2\nonumber\\
&+&\!\!\!\!\frac{2k^2(1-\mu)^2(1-\nu)H(x,y)}{(1-\lambda)(1-\mu\nu)\Phi\Psi(x-y)^2}\left[\frac{d x^2}{G(x)}-\frac{ dy^2}{G(y)}\right],
\eea
where
\bea
\Phi&=&1-\lambda\mu-\lambda\nu+\mu\nu\,,\qquad
\Psi=\mu-\lambda\nu+\mu\nu-\lambda\mu^2\,,\nn\\
\Xi&=&\mu+\lambda\nu-\mu\nu-\lambda\mu^2\,, \qquad G(x)=\left(1-x^2\right)(1+\mu x)(1+\nu x)\,,
\eea
also $H(x,y), J(x,y), F(x,y), \omega_\phi(x,y)$ and $\omega_\psi(x,y)$ are a bit messy functions that we refer the reader to \cite{Chen:2011jb} to see them. In this solution $x,y$ coordinates lie in the ranges $-1\leq x\leq 1$, $-\infty <y\leq -1$ where the infinity is at $x=y=-1$ and $0\leq \phi, \psi \leq 2\pi$. There are also four parameters $0\leq \nu \leq \mu \leq \lambda<1$ which are dimensionless and $k>0$ which has the dimension of length and sets the scale of the solution. 
Doing the same steps as (\ref{hmexp})-(\ref{k}), one can find the conical characteristic for the metric (\ref{ubr}) as
\be \label{ubrk}
\kappa_-=1\,, \qquad \kappa=\kappa_+=\frac{1+\mu}{1-\mu}\,\sqrt{\frac{(1-\lambda)(1+\nu)\Psi}{(1+\lambda)(1-\nu)\Xi}}\,.
\ee
By setting $\kappa_+=1$ one finds the constraint $\lambda=\frac{2\mu}{1+\mu^2}$ which resolves the conical singularity. Inserting this constraint to the parameters, one can recover the Pomeransky-Sen'kov (balanced) solution \cite{Pomeransky:2006bd}.  
The metric (\ref{ubr}) has two horizons (roots of $G(y)$), that the outer one is $y_+=-1/\mu$ and the inner is located at $y_-=-1/\nu$\,. The horizon area, Hawking temperature, mass and angular momenta of this solution is given in \cite{Chen:2011jb} as
\bea 
A_+&=&\frac{16\pi^2 k^3(\mu+\nu)(1-\mu)\Xi}{(1-\lambda)(1+\mu)}\bigg[\frac{2\lambda(1+\lambda)(1-\nu)}{(1-\mu\nu)^3\Phi\Psi}\bigg]^{\!\frac12}\,, \nn\\ T_+&=&\frac{(\mu-\nu)(1-\lambda)(1+\mu)}{8\pi k(\mu+\nu)(1-\mu)\Xi}\bigg[\frac{2(1-\mu\nu)\Phi\Psi}{\lambda(1+\lambda)(1-\nu)}\bigg]^{\!\frac12}\,,\nn\\
M&=&\frac{3\pi k^2\!\lambda(\mu+\nu)(1\!-\mu)\Phi}{2(1\!-\lambda)(1\!-\mu\nu)\Psi}\,,\qquad  J_{\phi}=\frac{2\pi k^3(\mu+\nu)(1-\mu)}{(1-\mu\nu)^{\frac32}}\bigg[\frac{2\nu\lambda(1+\lambda)\Xi}{(1-\lambda)\Phi\Psi}\bigg]^{\!\frac12}\,,\\
J_{\psi}&=&\frac{\pi k^3(\mu+\nu)(1\!-\mu)[2\nu(1\!-\!\lambda)(1\!-\mu)+(1\!-\nu)\Phi]}{(1-\lambda)^{\frac32}\,(1-\mu\nu)^{\frac32}\, \Psi^{\frac32}}\bigg[\frac{2\lambda(\lambda-\mu)(1+\lambda)(1-\lambda\mu)\Xi}{\Phi}\bigg]^{\!\frac12}\,,\nn\\
\Omega_\psi^{+}\!&=&\!\frac{1}{k(1\!-\!\mu)}\!\left[\frac{(\lambda\!-\!\mu)(1\!-\!\lambda)(1\!-\!\lambda\mu)(1\!-\!\mu\nu)\Psi}{2\lambda(1+\lambda)\Phi\Xi}\right]^{\!\!\frac 12}\!,\qquad \Omega_\phi^{+}\!=\!\frac{1+\mu}{k(\mu\!+\!\nu)}\!\left[\frac{\nu(1\!-\!\lambda)(1\!-\!\mu\nu)\Psi}{2\lambda(1+\lambda)\Phi\Xi}\right]^{\!\!\frac 12},\nn
\eea
where satisfy the Smarr relation $M=\frac{3}{2}\left[T_+ S_+ +\Omega_{\phi}^+ J_{\phi}+\Omega_{\psi}^+ J_{\psi}\right]$\, with $S=A/4$\,. We also computed the area, temperature and angular velocities of the inner horizon as 
\bea \label{ubrin}
A_-&=&\frac{16 \sqrt{2}\, \pi ^2 k^3 \nu (\lambda +1) (\mu -1)^2  (\mu +\nu )}{(\nu +1) \Psi }\bigg[\frac{\lambda  (\nu +1) \Xi }{(\lambda -1) \Phi  (\mu  \nu -1)^3}\bigg]^{\frac12}\,,\nn\\
T_-&=&\frac{(\nu +1)(-\mu+\nu) \Psi}{4 \sqrt{2}\, \pi  k (\lambda +1) (\mu -1)^2 (\mu +\nu)(\mu \nu -1)\nu}\bigg[\frac{(\lambda -1) \Phi  (\mu  \nu -1)^3}{\lambda  (\nu +1) \Xi }\bigg]^{\frac12}\,,\nn\\
\Omega_{\phi}^- &=&\frac{\lambda\,\mu^2-\lambda\,\nu^2+\mu\,\nu^2-
\mu}{k\left(1-\mu \right)\left(\mu+\nu\right)\left(1+\lambda
 \right)}\bigg[\frac{\Psi(\mu\nu-1) \left( {\lambda}^{2}-1 \right)}{2\,\nu\lambda\Phi\,\Xi}\bigg]^{\!\frac12}
\,,\nn\\
\Omega_{\psi}^-&=&\frac{\sqrt{\left(\lambda-\mu \right)  \left( \lambda\,\mu-1 \right)}}{k\,\Xi
 \left(\mu-1 \right)\left(1+\lambda \right)} \bigg[\frac{\Psi(\mu\nu-1) \left( {\lambda}^{2}-1 \right)}{2\lambda\Phi}\bigg]^{\!\frac12}\,.
 \eea
Considering (\ref{ubrk})-(\ref{ubrin}), it is easy to check that $T_+S_+=T_-S_-$ and the universality of area product in this case as
\be\label{uni2}
\kappa\,A_-\,A_+=64\,\pi^2J_{\phi}^2\,.
\ee
Similar to the previous cases, the conical character $\kappa$ appears in this relation. One can also check that thermodynamical quantities on the inner horizon for this solution satisfy the first law of thermodynamics and Smarr relation
\be\label{smarrubr}
dM=-T_- dS_-+{\Omega_{\phi}^-}\, dJ_{\phi}+{\Omega_{\psi}^-}\, dJ_{\psi}\,, \qquad M=\frac{3}{2}\left[-T_- S_-+{\Omega_{\phi}^-} J_{\phi}+{\Omega_{\psi}^-} J_{\psi}\right]\,,
\ee

\subsection{The dipole black ring}
A generalization of single rotating black ring which contains magnetic dipole charge is presented in \cite{dipole}. The solution is
\bea \label{dibr}
&ds^2&=-\frac{F(y)H(x)}{F(x)H(y)}\left[dt+R\sqrt{\lambda(\lambda-
\nu)\frac{1+\lambda}{1-\lambda}}\,\frac{1+y}{F(y)}d\psi\right]^2\\
&+&\frac{R^2F(x)H(x)H(y)^2}{(x-y)^2} \left[\frac{G(x)}{F(x)H(x)^3}d\phi^2+\frac{dx^2}{G(x)}
-\frac{dy^2}{G(y)}-\frac{G(y)}{F(y)H(y)^3}d\psi^2
\right],\nn
\eea
where the functions $F, G, H$ and the gauge field are
\bea
F(\xi)&=&1+\lambda\xi\,,\qquad G(\xi)=(1-\xi^2)(1+\nu\xi)\,,\nn\\
H(\xi)&=&1-\mu\xi\,, \qquad A_\phi=R\frac{1\!+\!x}{H(x)}\sqrt{3\mu(\mu\!+\!
\nu)\frac{1\!-\mu}{1\!+\mu}}+k\,,
\eea
which $k$ is a constant. The solution (\ref{dibr}) is characterized by four parameters $\mu, \nu, \lambda$ and $R$ that lie in the ranges $0\leq \mu<1$, $0<\nu \leq \lambda<1$ and $R>0$. The $x,y$ are also ring coordinates similar to the previous cases. In this solution $\lambda,\nu$ are related to the shape and rotation velocity, $R$ sets the scale of the ring and $\mu$ controls the dipole charge. The conical characteristic for this solution can be find according to (\ref{hmexp})-(\ref{k}) as
\be
\kappa_-=1\,, \qquad \kappa=\kappa_+=\frac {\nu+1}{\nu-1} \sqrt{\frac{(\mu+1)^3 (\lambda-1)}{(\mu-1)^3(\lambda+1)}}
\ee
The solution has an outer horizon at $y=-1/\nu$, and an inner one at $y=-\infty$\,. Horizon area, temperature, mass, angular momentum and the dipole charge for this black ring are \cite{dipole}
\bea
&&A_+=8\pi^2 R^3\frac{(1+\mu)^3(\mu+\nu)^{3/2}\sqrt{\lambda(1-
\lambda^2)}}{(1-\nu)^2(1+\nu)}\,,\nn\\
&&T_+=\frac{\nu(1+\nu)}{4\pi R(\mu+\nu)^{3/2}}\sqrt{\frac{1-
\lambda}{\lambda(1+\lambda)}}\,,\nn\\
&&M=\frac{3\pi R^2(1+\mu)^3}{4(1-
\nu)}\left[\lambda+\frac{\mu(1-\lambda)}{1+\mu}\right]\,,\nn\\
&&J^\psi=\frac{\pi R^3(1+\mu)^{9/2}}{2 (1-\nu)^2} \sqrt{\lambda(\lambda-\nu)(1+\lambda)}\,,\nn\\
&& q=\frac{R(1+\mu)(2\pi)^{1/3}}{(1-\nu)\sqrt{1-\mu}}\sqrt{\mu(\mu+\nu)(1-\lambda)}\,.
\eea
One can also calculate the horizon area, temperature, angular velocity and potential of the inner horizon as \cite{Castro:2012av}
\bea
&&A_-=8\pi^2R^3\frac{(1+\mu)^3\mu^{3/2}}{(1-\nu)^2}\sqrt{\lambda(\lambda-\nu)(1-\lambda^2)}\,,\nonumber\\
&&T_-=\frac{\nu}{4\pi R }\sqrt{1-\lambda\over \mu^3 (1+\lambda)(\lambda-\nu)}\,,\nn\\
&&\Omega^{\psi}_-=\frac{1-\nu}{R}\sqrt{\frac{\lambda}{(1+\mu)^{3}(1+\lambda)(\lambda-\nu)}}\,,\nn\\
&&\Phi_-=\frac{3 R(1+\mu)\pi^{2/3}}{2^{4/3}(1-\nu)} \sqrt{\frac{3(\mu+\nu)(1-\mu)(1-\lambda)}{\mu}}\,,
\eea
now it is easy to check that $T_+ S_+=T_-S_-$ and universality of the area product takes to the form (note the appearance of $\kappa$)
\be\label{uni}
\kappa\,A_-\,A_+=64\,\pi^2J\,q^3.
\ee
As well as the outer horizon, the first law of black holes inner mechanics and the Smarr relation are satisfied for this solution
\be\label{firstlaw}
dM=-T_-\frac{d A_-}{4} +({\Omega_-^\psi} \,d J^\psi +  \Phi_{-}\, dq)\,,\quad M=\frac{3}{2}\left[-T_- \,\frac{A_-}{4}+{\Omega_-^\psi} J^\psi\right] +\frac{1}{2} \Phi_{-}\, q\,.
\ee


\section{conclusions} \label{s4}
In this work we explored the area product for black solutions that contain conical singularity. We considered the charged C-metric, Black ring with rotating $S^2$, the unbalanced double rotating black ring and the dipole black ring. For all these solutions we observed that the area product of inner and outer horizons is mass independent which means that the universality of the area product is true for them. 

An interesting point for the solutions with conical singularity is the appearance of the conical characteristic ($\kappa$) in the universality relation as $\kappa A_+A_-=(8\pi)^2 N$, rather than $A_+A_-=(8\pi)^2 N$\, for the regular solutions, where $N$ is related to the quantized charges of the solutions. We observed this behavior for both black hole and black ring solutions. In the other word this property is true for the solutions, regardless of the topology of their horizons. We also computed the thermodynamical quantities on the inner horizon and checked that the first law of thermodynamics and the Smarr relation are satisfied for the solutions with conical singularity on the inner horizon as well as the outer horizon.


\section*{Acknowledgment}
I would like to thank M. M. Sheikh-Jabbari and Kamal Hajian for useful discussions.




\begin{thebibliography}{99}


\bibitem{Horowitz:1996fn} 
  G.~T.~Horowitz and A.~Strominger,
 ``Counting states of near extremal black holes,''
  Phys.\ Rev.\ Lett.\  {\bf 77}, 2368 (1996)
  [hep-th/9602051].   
 
\bibitem{Horowitz:1996ay} 
  G.~T.~Horowitz, J.~M.~Maldacena and A.~Strominger,
``Nonextremal black hole microstates and U duality,''
  Phys.\ Lett.\ B {\bf 383}, 151 (1996)
  [hep-th/9603109]. 
  
\bibitem{Halyo:1996xe} 
  E.~Halyo, B.~Kol, A.~Rajaraman and L.~Susskind,
  ``Counting Schwarzschild and charged black holes,''
  Phys.\ Lett.\ B {\bf 401}, 15 (1997)
  [hep-th/9609075].
  
  \bibitem{Larsen:1997ge} 
  F.~Larsen,
``A String model of black hole microstates,''
  Phys.\ Rev.\ D {\bf 56}, 1005 (1997)
  [hep-th/9702153].
  
\bibitem{Visser:2012zi} 
  M.~Visser,
 ``Quantization of area for event and Cauchy horizons of the Kerr-Newman black hole,''
  JHEP {\bf 1206}, 023 (2012)
  [arXiv:1204.3138 [gr-qc]].  
  
\bibitem{Cvetic:1997uw} 
  M.~Cvetic and F.~Larsen,
  ``General rotating black holes in string theory: Grey body factors and event horizons,''
  Phys.\ Rev.\ D {\bf 56}, 4994 (1997)
  [hep-th/9705192].
  
\bibitem{Galli:2011fq} 
  P.~Galli, T.~Ortin, J.~Perz and C.~S.~Shahbazi,
  ``Non-extremal black holes of N=2, d=4 supergravity,''
  JHEP {\bf 1107}, 041 (2011)
  [arXiv:1105.3311 [hep-th]].  

\bibitem{Cvetic:2010mn} 
  M.~Cvetic, G.~W.~Gibbons and C.~N.~Pope,
``Universal Area Product Formulae for Rotating and Charged Black Holes in Four and Higher Dimensions,''
  Phys.\ Rev.\ Lett.\  {\bf 106}, 121301 (2011)
  [arXiv:1011.0008 [hep-th]].  

\bibitem{Castro:2012av} 
  A.~Castro and M.~J.~Rodriguez,
``Universal properties and the first law of black hole inner mechanics,''
  Phys.\ Rev.\ D {\bf 86}, 024008 (2012)
  [arXiv:1204.1284 [hep-th]].
   
\bibitem{Ansorg:2009yi} 
  M.~Ansorg and J.~Hennig,
 ``The Inner Cauchy horizon of axisymmetric and stationary black holes with surrounding matter in Einstein-Maxwell theory,''
  Phys.\ Rev.\ Lett.\  {\bf 102}, 221102 (2009)
  [arXiv:0903.5405 [gr-qc]].
  
\bibitem{Ansorg:2010ru} 
  M.~Ansorg, J.~Hennig and C.~Cederbaum,
 ``Universal properties of distorted Kerr-Newman black holes,''
  Gen.\ Rel.\ Grav.\  {\bf 43}, 1205 (2011)
  [arXiv:1005.3128 [gr-qc]].  

\bibitem{Visser:2012wu} 
  M.~Visser,
 ``Area products for stationary black hole horizons,''
  Phys.\ Rev.\ D {\bf 88}, no. 4, 044014 (2013)
  [arXiv:1205.6814 [hep-th]].
      
\bibitem{Cvetic:2013eda} 
  M.~Cvetic, H.~Lu and C.~N.~Pope,
 ``Entropy-Product Rules for Charged Rotating Black Holes,''
  Phys.\ Rev.\ D {\bf 88}, 044046 (2013)
  [arXiv:1306.4522 [hep-th]].

\bibitem{Chen:2012mh} 
  B.~Chen, S.~x.~Liu and J.~j.~Zhang,
 ``Thermodynamics of Black Hole Horizons and Kerr/CFT Correspondence,'' JHEP {\bf 1211}, 017 (2012)
  [arXiv:1206.2015 [hep-th]].
  
  \bibitem{Chen:2012yd}
 B.~Chen and J.~-j.~Zhang,
  ``Holographic Descriptions of Black Rings,''
  JHEP {\bf 1211}, 022 (2012)
  [arXiv:1208.4413 [hep-th]].

  \bibitem{Castro:2013pqa} 
  A.~Castro, N.~Dehmami, G.~Giribet and D.~Kastor,
 ``On the Universality of Inner Black Hole Mechanics and Higher Curvature Gravity,''
  JHEP {\bf 1307}, 164 (2013)
  [arXiv:1304.1696 [hep-th]].

  \bibitem{Xu:2015mna} 
  W.~Xu, J.~Wang and X.~h.~Meng,
 ``Entropy bound of horizons for charged and rotating black holes,''
  Phys.\ Lett.\ B {\bf 746}, 53 (2015).
  
\bibitem{Debnath:2015tda} 
  U.~Debnath,
``Entropy bound of horizons for accelerating, rotating and charged Plebanski–Demianski black hole,''
  Annals Phys.\  {\bf 372}, 449 (2016)
  [arXiv:1507.00901 [gr-qc]].    
  
  \bibitem{Faraoni:2012je} 
  V.~Faraoni and A.~F.~Z.~Moreno,
  ``Are quantization rules for horizon areas universal?,''
  Phys.\ Rev.\ D {\bf 88}, no. 4, 044011 (2013)
  [arXiv:1208.3814 [hep-th]].
      
\bibitem{Wang:2013smb} 
  J.~Wang, W.~Xu and X.~H.~Meng,
  ``The 'universal property' of horizon entropy sum of black holes in four dimensional asymptotical (anti-)de-Sitter spacetime background,''
  JHEP {\bf 1401}, 031 (2014)
  [arXiv:1310.6811 [gr-qc]].
  
\bibitem{Pradhan:2015wnl} 
  P.~Pradhan,
 ``Area (or entropy) product formula for a regular black hole,''
  Gen.\ Rel.\ Grav.\  {\bf 48}, no. 2, 19 (2016)
  [arXiv:1512.06187 [gr-qc]].

 \bibitem{Anacleto:2013esa} 
  M.~A.~Anacleto, F.~A.~Brito and E.~Passos,
``Acoustic Black Holes and Universal Aspects of Area Products,''
  Phys.\ Lett.\ A {\bf 380}, 1105 (2016)
  [arXiv:1309.1486 [hep-th]].
       
\bibitem{Emparan:2001wn}
  R.~Emparan and H.~S.~Reall,
  ``A Rotating black ring solution in five-dimensions,''
  Phys.\ Rev.\ Lett.\  {\bf 88}, 101101 (2002)
  [hep-th/0110260].
  
  \bibitem{Emparan:2008eg}
 R.~Emparan and H.~S.~Reall,
 ``Black Holes in Higher Dimensions,''
 Living Rev.\ Rel.\  {\bf 11}, 6 (2008)
 [arXiv:0801.3471 [hep-th]].

\bibitem{Guica:2008mu}
  M.~Guica, T.~Hartman, W.~Song and A.~Strominger,
  ``The Kerr/CFT Correspondence,''
  Phys.\ Rev.\ D {\bf 80}, 124008 (2009)
  [arXiv:0809.4266 [hep-th]].
  
 \bibitem{Hartman:2008pb} 
  T.~Hartman, K.~Murata, T.~Nishioka and A.~Strominger,
 ``CFT Duals for Extreme Black Holes,''
  JHEP {\bf 0904}, 019 (2009)
  [arXiv:0811.4393 [hep-th]]. 

 \bibitem{Ghodsi:2013soa}
  A.~Ghodsi, H.~Golchin and M.~M.~Sheikh-Jabbari,
  ``Dual 2d CFT Identification of Extremal Black Rings from Holes,'' JHEP {\bf 1310}, 194 (2013)
  arXiv:1308.1478 [hep-th].

  \bibitem{Ghodsi:2014fta} 
  A.~Ghodsi, H.~Golchin and M.~M.~Sheikh-Jabbari,
 ``More on Five Dimensional EVH Black Rings,''  JHEP {\bf 1409}, 036 (2014)
  [arXiv:1407.7484 [hep-th]].
  
  \bibitem{Sadeghian:2015hja} 
  S.~Sadeghian and H.~Yavartanoo,
 ``Black rings in U(1)$^3$ supergravity and their dual 2d CFT,''  Class.\ Quant.\ Grav.\  {\bf 33}, no. 9, 095006 (2016) 
  [arXiv:1510.01209 [hep-th]].
      
\bibitem{Newman:1965my} 
  E.~T.~Newman, R.~Couch, K.~Chinnapared, A.~Exton, A.~Prakash and R.~Torrence,
 ``Metric of a Rotating, Charged Mass,''
  J.\ Math.\ Phys.\  {\bf 6}, 918 (1965).
  
\bibitem{Cvetic:1996kv} 
  M.~Cvetic and D.~Youm,
 ``Entropy of nonextreme charged rotating black holes in string theory,''
  Phys.\ Rev.\ D {\bf 54}, 2612 (1996)
  [hep-th/9603147].

\bibitem{Cvetic:1996xz} 
  M.~Cvetic and D.~Youm,
 ``General rotating five-dimensional black holes of toroidally compactified heterotic string,''
  Nucl.\ Phys.\ B {\bf 476}, 118 (1996)
  [hep-th/9603100].
   
\bibitem{Pomeransky:2006bd}
  A.~A.~Pomeransky and R.~A.~Sen'kov,
  ``Black ring with two angular momenta,''
  hep-th/0612005.

\bibitem{dipole}
  R.~Emparan,
 ``Rotating circular strings, and infinite nonuniqueness of black rings,''
  JHEP {\bf 0403}, 064 (2004)
  [hep-th/0402149].

\bibitem{kw}
W.~Kinnersley and M.~Walker, "Uniformly accelerating charged mass in general relativity,"
Phys. Rev. D {\bf 2} (1970) 1359.  

\bibitem{Hong:2004dm} 
  K.~Hong and E.~Teo,
 ``A New form of the rotating C-metric,''
  Class.\ Quant.\ Grav.\  {\bf 22}, 109 (2005)
  [gr-qc/0410002].

\bibitem{Griffiths:2005qp} 
  J.~B.~Griffiths and J.~Podolsky,
``A New look at the Plebanski-Demianski family of solutions,''
  Int.\ J.\ Mod.\ Phys.\ D {\bf 15}, 335 (2006)
  [gr-qc/0511091].

\bibitem{Astorino:2016ybm} 
  M.~Astorino,
 ``Thermodynamics of Regular Accelerating Black Holes,''
  Phys.\ Rev.\ D {\bf 95}, no. 6, 064007 (2017)
  [arXiv:1612.04387 [gr-qc]].
  
\bibitem{Figueras:2005zp}
  P.~Figueras,
  ``A Black ring with a rotating 2-sphere,''
  JHEP {\bf 0507}, 039 (2005)
  [hep-th/0505244].

\bibitem{Chen:2011jb}
  Y.~Chen, K.~Hong and E.~Teo,
  ``Unbalanced Pomeransky-Sen'kov black ring,''
  Phys.\ Rev.\ D {\bf 84}, 084030 (2011)
  [arXiv:1108.1849 [hep-th]].
   
  \end{thebibliography}
 \end{document}